\documentclass[twocolumn,superscriptaddress,
amsmath,amssymb,
longbibliography,reprint]{revtex4-2}

\usepackage{amssymb,amsmath}
\DeclareMathAlphabet\mathbfcal{OMS}{cmsy}{b}{n}
\usepackage{cmap}
\usepackage{graphicx} 
\usepackage{bm}
\usepackage{color}
\usepackage{blindtext}
\usepackage{hyperref}
\usepackage{listings}
\usepackage{booktabs}
\usepackage{multirow}
\usepackage{amsmath}
\usepackage{mathrsfs}
\usepackage{braket}
\usepackage{mathtools}

\usepackage{dsfont}
\usepackage{tabularx}

\begin{document}


\title{Bulk-to-surface coupling of spatiotemporal vortices}

\author{Francisco J. Rodr{\'\i}guez-Fortu{\~n}o}
\affiliation{King's College London, Department of Physics and London Centre for Nanotechnology, London, United Kingdom}

\author{Michela F. Picardi}
\affiliation{ICFO --- The Institute of Photonics Sciences,
The Barcelona Institute of Science and Technology, Castelldefels, Barcelona 08860, Spain}

\author{Konstantin Y. Bliokh}
\affiliation{Donostia International Physics Center (DIPC), Donostia-San Sebasti\'{a}n 20018, Spain}
\affiliation{IKERBASQUE, Basque Foundation for Science, Bilbao 48009, Spain}
\affiliation{Centre of Excellence ENSEMBLE3 Sp. z o.o., 01-919 Warsaw, Poland}


\begin{abstract}
We consider the coupling of bulk space-time-structured waves, such as spatiotemporal vortex pulses (STVPs), to surface waves, such as surface plasmon-polaritons (SPPs). For resonant coupling that preserves the frequency and tangent wavevector components, it is challenging to transfer the space-time wave structure because of the difference between the bulk and surface wave dispersions. We describe three mechanisms allowing for the bulk-to-surface conversion of STVPs: (i) a suitable tilt of the pulse spectrum in the $(\omega,{\bf k})$ space; (ii) confinement of the incident pulse in the direction orthogonal to the vortex; and (iii) losses in the surface waves. We supply general considerations with numerical simulations of the STVP-to-SPP coupling at a dielectric-prism/metal-film interface.   
\end{abstract}

\maketitle


Following remarkable progress in studies and applications involving structured (i.e., inhomogeneous) optical, acoustic, and other types of waves \cite{Andrews2008, Rubinsztein2016, Bliokh2023JO}, there is a current burst of interest in {\it spatiotemporal} structured waves, which are inhomogeneous in both space and time \cite{Yessenov2022, Shen2023}. Such waves are essentially {\it polychromatic}, which makes their consideration more {intricate}. In particular, the role of wave dispersion $\omega ({\bf k})$ ($\omega$ is the frequency, ${\bf k}$ is the wavevector) increases considerably. 

The generation and propagation of various spatiotemporal structured waves has been well explored by now. An important question, crucial for information-transfer applications, is: how can one transfer the wave structure between different optical elements, media, or even different kinds of wave? Here we consider this question in the context of coupling of bulk spatiotemporal structured waves into 2D surface or planar-guided modes. 

The Fourier amplitudes of spatiotemporal $(t,{\bf r})$-structured waves are accordingly structured in the $(\omega, {\bf k})$ domain. Importantly, this domain is constrained by dispersion relations, which are different for bulk and surface modes: $\omega ({\bf k})$ and $\omega_s ({\bf k}_s)$. Furthermore, resonant coupling to surface or guided waves is typically determined by phase-matching conditions preserving frequencies and tangent wavevector components. Therefore, maintaining the $(\omega, {\bf k})$-space structure simultaneously with the phase-matching conditions and different dispersion relations offers a challenging problem.


Here we address this problem by considering an important class of spatiotemporal structured waves: {\it spatiotemporal vortex pulses} (STVPs), which were recently explored both theoretically \cite{Sukhorukov2005, Bliokh2012PRA, Dror2011PD, Bliokh2021PRL, Hancock2021PRL,  Wang2021O, Mazanov2021, Porras2023OL_II, Bliokh2023PRA, Porras2024JO, Bekshaev2024APL, Vo2024JO} and experimentally \cite{Jhajj2016, Chong2020NP, Hancock2019O, Zang2022NP, Cao2023AP, Liu2024NC, Huang2024SA, Zhang2023NC, Ge2023PRL, Che2024PRL, Martin-Hernandez2025NP}. 
While well-known monochromatic vortex beams are 3D structures, carrying screw wavefront dislocations \cite{Allen_book, Torres2011}, STVPs can appear as moving 2D structures with edge wavefront dislocations. This makes them perfectly suited for surface waves (e.g., water waves \cite{Che2024PRL}) or planar waveguides (e.g., acoustic \cite{Zhang2023NC}). Yet, the generation of STVPs in 2D optical systems, such as surface plasmon-polaritons (SPPs) \cite{Maier_book, Zayats2005}, is still an open challenge. 

\begin{figure}[t!]
\centering
\includegraphics[width=0.9\linewidth]{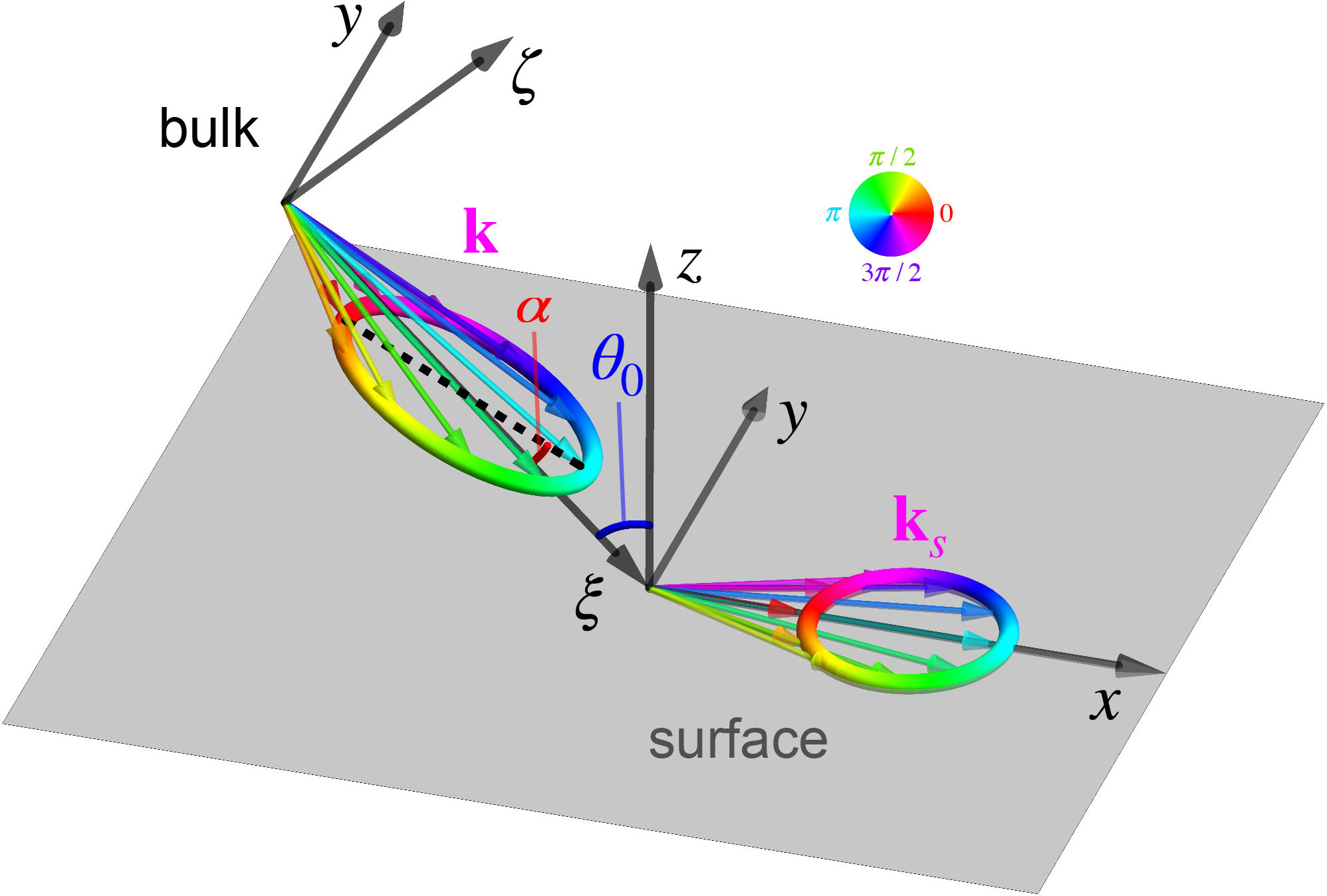}
\caption{Schematics of a spatiotemporal vortex pulse (STVP) with an elliptical (Bessel-type) plane-wave spectrum incident from bulk and exciting a surface-wave STVP. The coordinate frame $(\xi,y,\zeta)$ is rotated with respect to the $(x,y,z)$ frame by an angle $\pi/2 - \theta_0$ about the $y$-axis ($\theta_0$ is the angle of incidence), and $\alpha$ is an additional tilt of the plane of the incident pulse with respect to the {$(\xi,y)$} plane. Plane waves in the spectra are marked by colors corresponding to the azimuthal-angle parameter $\phi \in (0, 2\pi)$. \label{Fig1}}
\end{figure}

In this paper, we consider the resonant coupling and transformation of an obliquely incident bulk STVP into a 2D surface-wave (or planar-waveguide) STVP, Fig.~\ref{Fig1}. We assume a linear dispersion for bulk waves, an arbitrary dispersion for surface waves, and the simplest phase-matching conditions. For numerical simulations, we take an example of the SPP excitation via total internal reflection at the bottom of a glass prism (the Otto-Kretschmann configuration). We show that to preserve the spatiotemporal structure of the pulse, one needs to obey one of the three conditions: 

(i) The spectrum of the incident pulse is appropriately {\it tilted} in the $(\omega,{\bf k})$ space;

(ii) The incident pulse is sufficiently {\it confined} in the direction orthogonal to the vortex plane; 

(iii) The surface waves have sufficient {\it losses}.
\newline
The first condition addresses the difference between the bulk and surface wave dispersions, and provides the best matching between the Fourier spectra of these waves. 

Figure~\ref{Fig1} depicts schematics of the problem.
We aim at exciting a paraxial surface-wave STVP, which propagates in the $z=0$ plane along the $x$-axis. The simplest Bessel type of such STVP \cite{Bliokh2021PRL,Bliokh2023PRA} is described by the plane-wave spectrum distributed over an ellipse in the $(k_x,k_y)$ plane:
\begin{equation}
{\bf k}_s = (k_{s0} +\Delta_x \cos\phi, \Delta_y \sin\phi) \,.
\label{s_STVP}    
\end{equation}
Here, ${\bf k}_{s0}=(k_{s0},0)$ is the central (mean) wavevector, $\Delta_{x,y}$ are the spectral widths of the pulse along the $x$ and $y$ axes, and $\phi \in (0,2\pi)$ is the azimuthal-angle parameter along the elliptical curve in the wavevector space. 
In the paraxial approximation, we assume $\Delta_{x,y} \ll k_{s0}$, and then the frequencies, corresponding to the spectrum (\ref{s_STVP}) and dispersion relation $\omega_s (k_s)$, are
\begin{equation}
\omega_s (k_s) \simeq \omega_{s0} + \frac{\partial \omega_{s0}}{\partial k_{s0}} \Delta_x \cos\phi \,,
\label{s_STVP_omega}    
\end{equation}
where $\omega_{s0} = \omega_{s} (k_{s0})$.


The vortex structure is determined by the phase factor $e^{i\ell\phi}$ ($\ell \in \mathbb{Z}$ is the topological vortex charge) in the spectral plane-wave amplitudes. Notably, the topological charge does not affect the considerations below, because the wavevectors and frequencies of Bessel-type vortices are independent of $\ell$.

We assume that the surface-wave STVP is excited by a paraxial bulk-wave STVP incident at an angle $\theta_0$, as shown in Fig.~\ref{Fig1}. The plane-wave spectrum of the incident pulse can be written as 
\begin{equation}
{\bf k} = {\bf k}_0 + {\bm \delta} = (k_{0}\sin\theta_0 +\delta_x, \delta_y, - k_{0}\cos\theta_0 +\delta_z) \,,
\label{b_STVP}    
\end{equation}
where ${\bf k}_0$ is the central wavevector and ${\bm \delta}$ describes small deflections from it ($\delta \ll k_0$). For a linear bulk-wave dispersion $\omega = \tilde{c}k$, the corresponding frequencies are
\begin{equation}
\omega (k) \simeq \omega_0 + \tilde{c}\, (\delta_x \sin\theta_0 - \delta_z \cos\theta_0) \,,
\label{b_STVP_omega}    
\end{equation}
where $\omega_0 = \tilde{c} k_0$, and $\tilde{c}$ is the speed of the bulk wave.

Next, we assume that the system is invariant under space-time translations along the $(x,y,t)$ dimensions.  
This is the case, e.g., for the excitation of SPPs via total internal reflection at the bottom of a glass prism parallel to the metal surface  \cite{Maier_book, Zayats2005}. 
Then, the frequencies and tangent wave-vector components in the bulk and surface pulses have to match each other: 
\begin{equation}
\omega_s = \omega\,,\quad k_{s\,x} = k_x\,, \quad k_{s\,y} = k_y\,.
\label{matching}    
\end{equation}

Substituting Eqs.~(\ref{s_STVP}), (\ref{s_STVP_omega}), (\ref{b_STVP}), and (\ref{b_STVP_omega}) into the matching conditions (\ref{matching}), we obtain the necessary conditions for the bulk pulse to couple to the surface one:
\begin{align}
\label{matching_0}
\omega_0  = \omega_{s0}  \,, \quad k_0 \sin\theta_0 = k_{s0} \, 
\end{align}
\begin{align}
\delta_x & = \Delta_x\cos\phi \,, \quad
\delta_y = \Delta_y \sin\phi \,, \nonumber \\
\delta_z & = \Delta_x \cos\phi \left( \frac{\sin\theta_0}{\cos\theta_0} - \frac{1}{\tilde{c} \cos\theta_0} \frac{\partial \omega_{s0}}{\partial k_{s0}} \right).
\label{matching_1}
\end{align}
Equations (\ref{matching_0}) are the standard phase-matching conditions in the central-plane-wave approximation, while Eqs.~(\ref{matching_1}) take into account deflections of plane waves in the spectra from the central waves. 

Importantly, the last condition (\ref{matching_1}) involves the surface-wave {\it dispersion}. For non-dispersive surface waves with ${\partial \omega_{s0}}/{\partial k_{s0}} = \omega_{s0}/ k_{s0} = \tilde{c} / \sin\theta_0$, Eqs.~(\ref{matching_1}) result in $\delta_z \sin\theta_0 = - \delta_x \cos\theta_0$. In this case, the incident-pulse spectrum (\ref{b_STVP}) is an ellipse lying in the plane containing the central wavevector ${\bf k}_0$. In other words, such pulse has a purely transverse angular momentum, associated with the vortex \cite{Bliokh2021PRL, Hancock2021PRL,  Mazanov2021, Wang2021O, Bliokh2023PRA, Porras2024JO, Vo2024JO, Jhajj2016, Chong2020NP, Hancock2019O, Zang2022NP, Cao2023AP, Liu2024NC, Huang2024SA, Zhang2023NC, Ge2023PRL, Che2024PRL, Wan2023eL, Chen2023ACS}. However, for dispersive surface waves, ${\partial \omega_{s0}}/{\partial k_{s0}} \neq \omega_{s0}/ k_{s0}$, the plane-wave spectrum (\ref{b_STVP}) and (\ref{matching_1}) is an ellipse {\it tilted} with respect to the central wavevector ${\bf k}_0$, and such pulse has a tilted vortex-induced angular momentum \cite{Bliokh2012PRA, Zang2022NP, Wang2021O, Porras2023OL_II}. 

To describe such a tilted STVP, 
%
\label{delta_z}
%
it is instructive to introduce the  coordinate frame $(\xi,y,\zeta)$ rotated by the angle $\pi/2 - \theta_0$ about the $y$-axis, Fig.~\ref{Fig1}. The central wavevector ${\bf k}_0$ of the incident pulse becomes aligned with the $\xi$-axis, whereas the deflection ${\bm \delta}$, Eq.~(\ref{matching_1}), acquires components
\begin{align}
\delta_\xi & = -\delta_z \cos\theta_0 + \delta_x \sin\theta_0 
 = \Delta_x \frac{\cos\phi}{\sin\theta_0} \frac{v_{s}}{u_{s}} \,, \nonumber \\
\delta_\zeta & = \delta_z \sin\theta_0 + \delta_x \cos\theta_0 
 = \Delta_x \frac{\cos\phi}{\cos\theta_0} \!\left( 1 - \frac{v_{s}}{u_{s}} \right).
\label{matching_2}
\end{align}
Here we introduced the group and phase velocities of the surface wave: $v_{s} = {\partial \omega_{s0}}/{\partial k_{s0}}$ and $u_{s} = {\omega_{s0}}/{k_{s0}}$.
Then, the {tilt} angle $\alpha$ of the elliptical spectrum of the incident STVP in the $(\xi,\zeta)$-coordinates is determined by
\begin{equation}
\tan \alpha = \frac{\delta_\zeta}{\delta_\xi} = \left( \frac{u_s}{v_s} -1 \right) \tan\theta_0 \,.
\label{tilt}    
\end{equation}
For SPPs, $u_s > v_s$, and, hence, $\alpha > 0$, as shown in Fig.~\ref{Fig1}.



\begin{figure}[t!]
\centering
\includegraphics[width=0.8\linewidth]{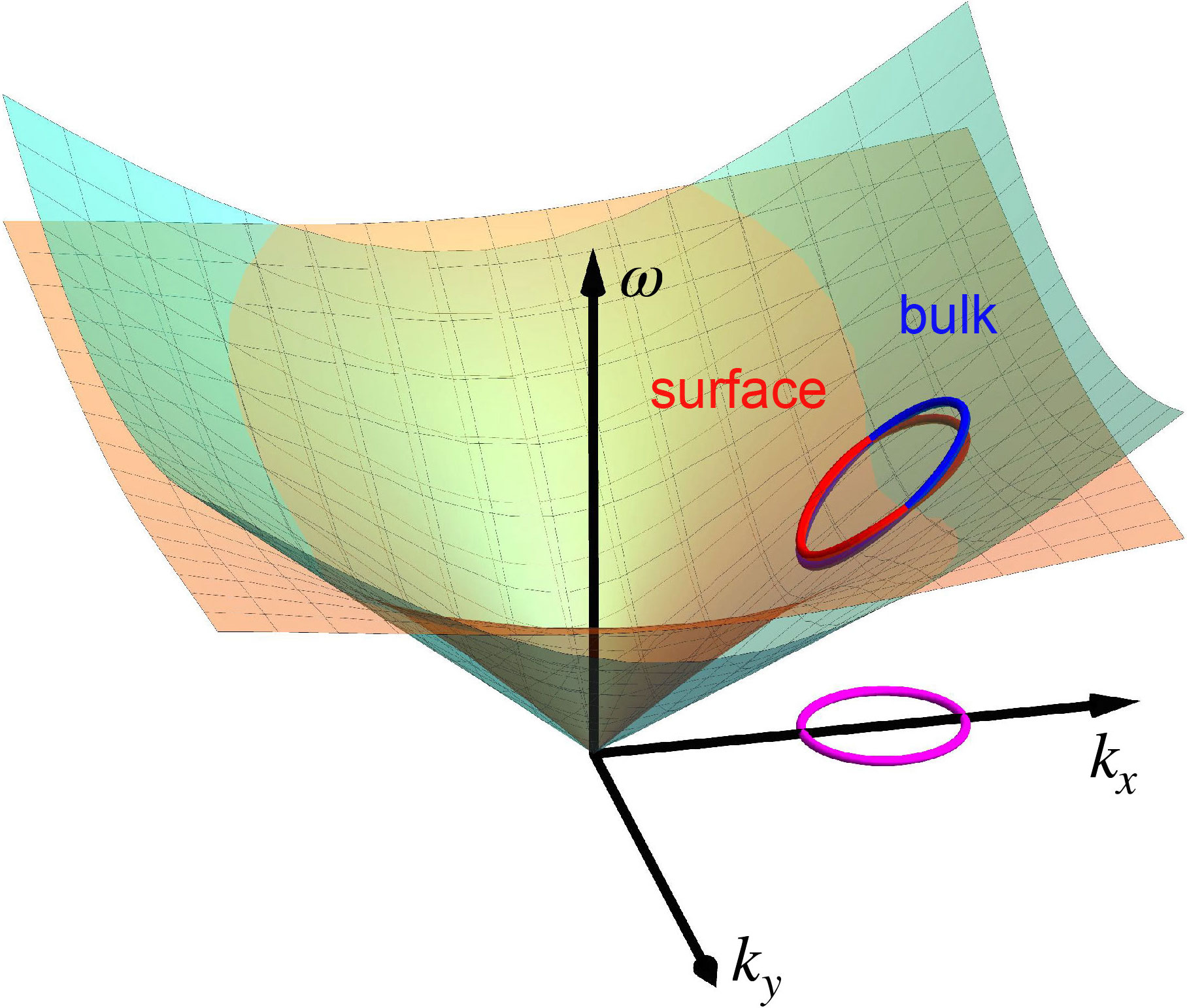}
\caption{Dispersion sheets for the surface (pink) and bulk (cyan) pulses (without the tilt $\alpha$ shown in Fig.~\ref{Fig1}), Eqs.~(\ref{dispersion_inc}) and (\ref{dispersion_SPP}). The blue and red curves depict the surface-STVP and bulk-STVP spectra corresponding to the same magenta ellipse in the $(k_x,k_y)$ plane. These curves intersect only in two points, but can be made overlapping by introducing finite thickness either to the incident pulse dispersion (transverse $\zeta$-confinement), or to the surface-wave dispersion (losses), Eqs.~(\ref{thickness}) and (\ref{losses}). \label{Fig2}}
\end{figure}

Incident STVP with a tilted spectrum, Eqs.~(\ref{matching_2}) and (\ref{tilt}), is the best option for the resonant coupling under consideration. However, many experiments deal with non-tilted STVPs with purely transverse angular momentum, and therefore we explore the possibilities of phase matching between bulk and surface STVPs without this additional tilt. We now assume $\alpha = \delta_\zeta = 0$ in the incident pulse, which yields $k_z = - k_x \cot\theta_0$ and 
\begin{equation}
\omega = \tilde{c}k = \tilde{c}\sqrt{k_x^2 + k_y^2 + k_z^2} = \tilde{c} \sqrt{\frac{k_x^2}{\sin^2\!\theta_0} + k_y^2} \,.
\label{dispersion_inc}    
\end{equation}
Figure~\ref{Fig2} shows this effective dispersion $\omega(k_x,k_y)$ together with the surface-wave dispersion, $\omega_s (k_{s\, x},k_{s\, y})$, where we used SPPs on the surface of a Drude-model metal (assuming free space above the surface) \cite{Maier_book, Zayats2005}: 
\begin{equation}
\omega_s = \sqrt{{c}^2 k_s^2 +\omega_p^2/2 -\sqrt{{c}^4 k_s^4 + \omega_p^4/4}} \,.
\label{dispersion_SPP}    
\end{equation}
Here, $k_s^2 = k^2_{s\, x}+k^2_{s\, y}$, $\omega_p$ is the electron plasma frequency in the metal, and $c$ is the speed of light in free space.

Let the phase matching (\ref{matching_0}) for the central wavevectors and frequencies be satisfied, and we deal with an elliptical Bessel-type spectrum in the $(k_x,k_y)$ plane (purple ellipse in Fig.~\ref{Fig2}). Then, this spectrum corresponds to different curves on the bulk-wave and surface-wave dispersion surfaces (blue and red curves in Fig.~\ref{Fig2}), which intersect only at two points. Thus, because of the difference between group velocities of the bulk and surface waves, the phase matching (\ref{matching}) is satisfied only for two plane waves in the pulse spectrum.

Fortunately, infinitely thin elliptical spectra, as well as infinitely thin dispersion sheets are idealized theoretical concepts. Indeed, an infinitely thin elliptical spectrum of the incident pulse in Fig.~\ref{Fig1} corresponds to a cylindrical pulse infinitely extended along the $\zeta$-axis (for a non-tilted pulse). To deal with physical pulses {\it confined} along the $\zeta$-axis, we introduce a Gaussian-like distribution $\exp(-D^2 k^2_\zeta/2)$ in the plane-wave spectrum, where $D$ is the $\zeta$-thickness of the pulse. Such distribution provides an effective `thickness' to the blue curve in Fig.~\ref{Fig2}, so that it can now overlap with the red curve. The condition for this can be found from the second Eq.~(\ref{matching_2}):  
\begin{align}
D \sim \frac{1}{\delta_\zeta} \lesssim \frac{\cos\theta_0}{\Delta_x} \!\left( 1 - \frac{v_{s}}{u_{s}} \right)^{-1}.
\label{thickness}
\end{align}
Under this condition, the Gaussian-distributed plane waves in the incident STVP include a tilted ellipse required for optimal phase matching (\ref{matching_2}) and excitation of the surface STVP.

\begin{figure*}[t!]
\centering
\includegraphics[width=0.9\linewidth]{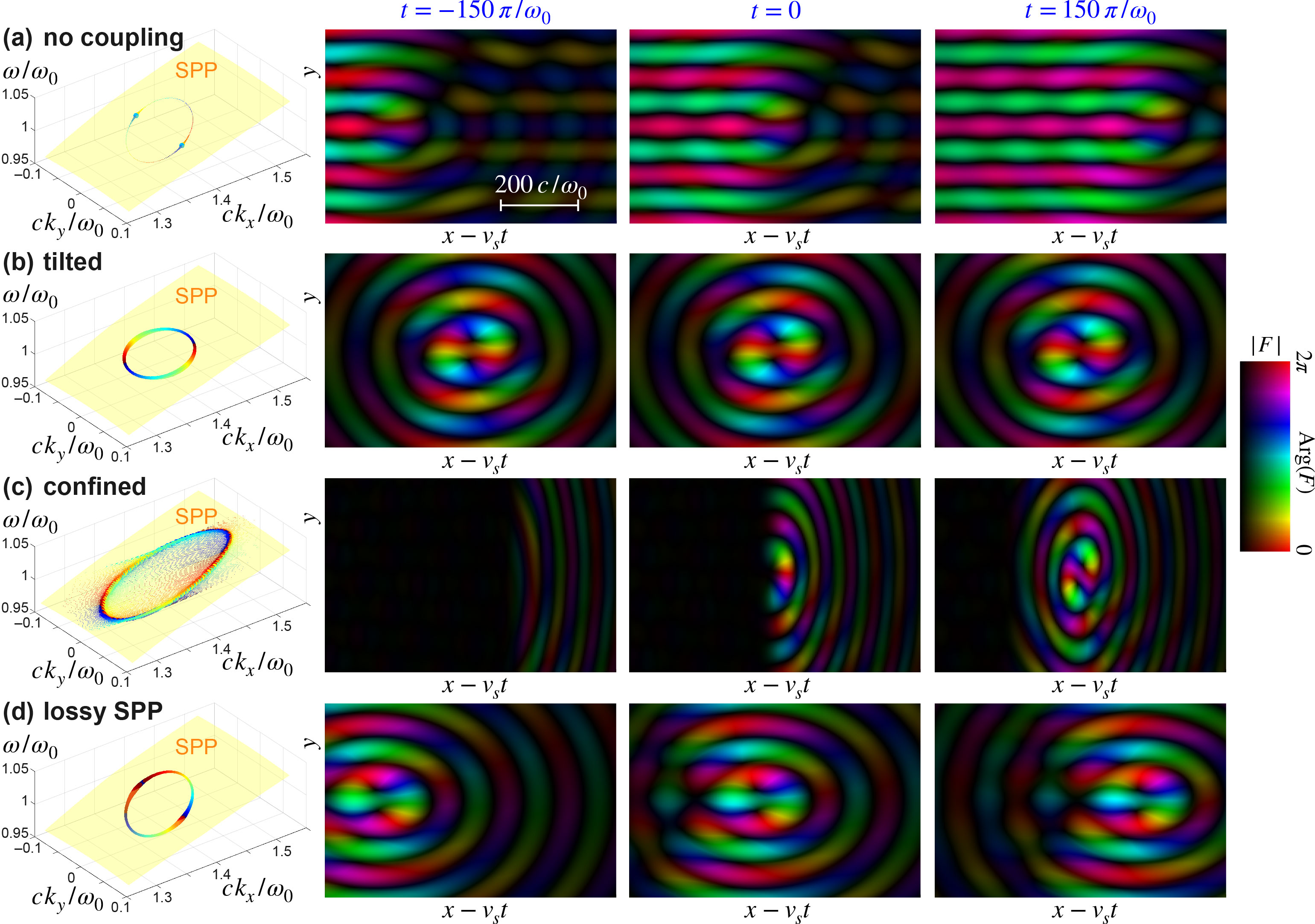}
\caption{Numerical simulations of the optical STVP to SPP coupling (see explanations in the text). The SPP field is presented by the normal electric-field component $E_{sz}$ at the metal-air interface, without the central-wave propagation factor: $F=E_{sz}e^{-ik_{0x}x}$. It is 
plotted at different moments of time $t$ in the frame $(x-v_st,y)$ moving with the SPP group velocity. The intensity and phase are encoded by the brightness and colors. (a) A ring-spectrum STVP without tilt ($\alpha=0$) couples to SPPs in two points (see Fig.~\ref{Fig2}) and does not transfer the vortex. (b) An optimized STVP with tilted spectrum, $\alpha=1.0808~\text{rad}$. (c) Transversely confined (non-tilted) STVP with $D=12.6\,c/\omega_0$. (d) Same STVP as in (a) but coupled to lossy SPPs with the collision frequency $\gamma = 10^{14}~\text{rad/s}$.   
\label{Fig3}}
\end{figure*}

Finally, there is another way to provide an overlap between the dispersion characteristics of bulk and surface STVPs in Fig.~\ref{Fig2}. Namely, even for an infinitely thin incident-STVP spectrum, {\it losses} in the surface waves effectively provide a finite thickness to their dispersion sheet, and hence to the red curve in Fig.~\ref{Fig2}. This thickness is $\delta\omega_s \sim |{\rm Im}\, \omega_s |$, where the small imaginary part ${\rm Im}\,\omega_s$ describes the losses.
In particular, such losses ({originating from absorption or leakage}) are inevitable in SPPs. 
Overlap between the (infinitely-thin) bulk and surface pulse dispersions in Fig.~\ref{Fig2} is achieved when
\begin{align}
\delta\omega_s \gtrsim \frac{\tilde{c}}{\sin\theta_0}\Delta_x - \frac{\partial \omega_{s0}}{\partial k_{s0}} \Delta_x 
= \left( u_s - v_s \right) \Delta_x\,.
\label{losses}
\end{align}
%



To illustrate different mechanisms of the bulk-to-surface coupling of STVPs, we performed numerical simulations shown in Fig.~\ref{Fig3}.
We modeled an optical STVP ($\omega_0 \simeq 7.54\times 10^{15}~\text{rad/s}$, $\Delta_x = \Delta_y=0.05\, \omega_0/c$, and $\ell=2$) with TM polarization propagating through a high-index glass (refractive index $n=1.7$) at $z>0$ and impinging onto a thin slab of gold (thickness $d\simeq 1.9\, c/\omega_0$), modeled via a lossless Drude model ($\omega_p \simeq  13.26\times10^{15}~\text{rad/s}$), with air on its other side ($z<-d)$. For the resonant incidence angle $\theta_0 \simeq 0.9488~\text{rad}$, the bulk wave couples into SPPs at the metal-air interface (the Otto-Kretschmann configuration \cite{Maier_book, Zayats2005}). We modeled the incident vortex via a finite array of its Fourier (plane-wave) components. Applying the appropriate Fresnel transmission coefficients $t_p(k_x,k_y,\omega)$ to each of the bulk plane waves, we obtained the Fourier spectrum of the SPP field. These SPP spectral components are shown in the left column of Fig.~\ref{Fig3} as dots, with the amplitudes and phases encoded by size and color, respectively. Interference of these components produces the SPP field distributions in real space.  Figure~\ref{Fig3}(a) shows a non-tilted ring spectrum with two-point coupling (see Fig.~\ref{Fig2}), which destroys the vortex and produces a two-wave-like SPP interference. Figures~\ref{Fig3}(b-d) illustrate the three coupling mechanisms (tilt, confinement, losses), clearly demonstrating the excitation of spatiotemporal SPP vortices.

{Note that Figs.~\ref{Fig3}(a,b,d) involve 1D ring spectra in the $(\omega,\mathbf{k})$ space, which correspond to cylindrical STVPs with infinite $\zeta$-thickness in real space. Such STVPs always intersect the slab, continuously producing surface vortices at $t$-varying positions. In contrast, Fig.~\ref{Fig3}(c) involves a disk-like $\zeta$-confined STVP exciting the pulse only when it reaches the slab.}
{Note also that propagation speed of the SPP pulse in Fig.~\ref{Fig3}(d) is faster than the SPP group velocity $v_s$; it is determined by the slope of the effective bulk-wave dispersion (\ref{dispersion_inc}) that fits within the loss-extended SPP resonance.} 


To conclude, we have described the main mechanisms of the resonant coupling of space-time-structured waves from bulk to surface modes. 
Specifically, we have examined the case of bulk STVPs coupled to SPPs, assuming conserved frequency and tangent wavevector components. 
While transverse confinement of the pulse and SPP losses facilitate  transfer of the space-time wave structure, a suitable rotation of the bulk pulse spectrum in the $(\omega,{\bf k})$ space allows for near-perfect adjustment between the bulk and surface wave dispersions. Other mechanisms, not considered in this work, might include broadband couplers, such as a single slit that removes conservation of $k_x$ \cite{Gorodetski2012PRL}, active media \cite{MacDonald2009NP}, and nonlinear effects \cite{Bliokh2009PRA}. 


We acknowledge support from Marie Sk\l{}odowska-Curie COFUND Programme of the European Commission (HORIZON-MSCA-2022-COFUND-101126600-SmartBRAIN3), 
ENSEMBLE3 Project carried out within the 
International Research Agendas Programme (IRAP) of the Foundation for Polish Science co-financed by the European Union under the European Regional Development Fund (MAB/2020/14), 
Teaming Horizon 2020 programme of the European Commission (GA No. 857543), 
Minister of Science and Higher Education ``Support for the activities of Centers of Excellence established in Poland under the Horizon 2020 program'' (MEiN/2023/DIR/3797), and 
{the UK EPSRC project META4D EP/Y015673.}







\bibliography{References_STVP}


\end{document}